# Resonant properties of dipole skyrmions in amorphous Fe/Gd multilayers

S. A. Montoya[1,2], S. Couture[1,2], J. J. Chess[3], J. C. T Lee[3,6], N. Kent[5,6], M.-Y. Im[6,7],
S. D. Kevan[3,6], P. Fischer[4,5], B. J. McMorran[3], S. Roy[6], V. Lomakin[1,2], and E.E. Fullerton[1,2] *

[1]*Center for Memory and Recording Research, University of California, San Diego, La Jolla, CA 92093, USA*
[2]*Department of Electrical and Computer Engineering, University of California, San Diego, La Jolla, CA 92093, USA*
[3]*Department of Physics, University of Oregon, Eugene OR 97401, USA*
[4]*Materials Sciences Division, Lawrence Berkeley National Laboratory, Berkeley CA 94720, USA*
[5]*Physics Department, University of California, Santa Cruz, CA 94056, US*
[6]*Center for X-ray Optics, Lawrence Berkeley National Laboratory, Berkeley, California 94720, USA*
[7]*Department of Emerging Materials Science, Daegu Gyeongbuk Institute of Science and Technology, Daegu, Korea*

Dated: May 3, 2017

## Abstract

The dynamic response of dipole skyrmions in Fe/Gd multilayer films is investigated by ferromagnetic resonance measurements and compared to micromagnetic simulations. We detail thickness and temperature dependent studies of the observed modes as well as the effects of magnetic field history on the resonant spectra. Correlation between the modes and the magnetic phase maps constructed from real-space imaging and scattering patterns allows us to conclude the resonant modes arise from local topological features such as dipole skyrmions but does not depend on the collective response of a closed packed lattice of these chiral textures. Using, micromagnetic modeling, we are able to quantitatively reproduce our experimental observations which suggests the existence of localized spin-wave modes that are dependent on the helicity of the dipole skyrmion. We identify four localized spin wave excitations for the skyrmions that are excited under either in-plane or out-of-plane *r.f.* fields. Lastly we show that dipole skyrmions and non-chiral bubble domains exhibit qualitatively different localized spin wave modes.

## I.    Introduction

Skyrmions in thin films are cylindrical-like magnetic domains with non-trivial topologies

---

* Corresponding author: efullerton@ucsd.edu

[1-4] that have been suggested as a building block for next generation mobile technologies [5-7]. These magnetic textures are usually stabilized by Dzyaloshinskii–Moriya interactions (DMI) in non-centrosymmetric bulk magnets [8-10] and thin films [12-15], but can also form under the competition of dipole and domain wall energy in magnetic thin films with perpendicular magnetic anisotropy [16-19]. Recently, it was shown that Bloch-type skyrmions stabilized by DMI exhibit local spin wave interactions that are independent of the class of material in which they form [20]. Insulating, semiconducting and metallic systems all exhibit similar spin-wave excitations that are solely dependent on the resonant mode and perturbation geometry [20, 21]. The Bloch-type DMI skyrmion dynamics under an in-plane microwave field is numerically predicted to result in a clockwise or counter-clockwise rotation of the spins inside the skyrmion core [22]; whereas an out-of-plane or perpendicular *r.f.* field perturbation results in the skyrmion core expansion and contraction, termed a breathing mode [22-24]. Comparable localized spin-wave dynamics have also been suggested for Néel-type DMI skyrmions under microwave fields [25]. The question remains whether skyrmions stabilized by mechanisms without the presence of DMI should exhibit similar spin wave modes determined by the topology.

Here we report the spectrum of resonant modes that result from magnetic domains ranging from stripes, magnetic bubbles and dipole skyrmions in amorphous Fe/Gd multilayers [17, 18]. The formation of dipole-stabilized biskyrmions and skyrmions in this class of material was previously reported in Refs. 17 and 18, respectively. These results highlight the materials properties, field ranges and temperature ranges where the skyrmion phase can be formed. By tuning the magnetization, uniaxial anisotropy, exchange interaction and film thickness we can stabilize isolated or closed-packed lattices of dipole skyrmions under the application of a magnetic field, which are sub-100-nm is size [18]. Unlike in DMI materials, the dipole skyrmion phase consists of an equal population of chiral domains with two possible helicities. Since the Bloch-line continuously wraps around the magnetic texture, we define it as a dipole skyrmion that possesses a winding number S = 1 [2-4]. If the Bloch-line wraps itself around the cylindrical-like domain in a clockwise direction it has helicity $\gamma = -\pi/2$; conversely, a Bloch-line wall that wraps counter-clockwise has helicity $\gamma = +\pi/2$ [2]. Figure 1a shows a schematic dipole skyrmion with $\gamma = +\pi/2$ helicity. These magnetic domains are primarily experimentally identified by resonant soft x-ray scattering patterns performed at the Gd $M_5$ (1198 eV) absorption edge and Fe $L_3$ (706 eV) absorption edge and real space images obtained by means of Lorentz TEM and full-field transmission resonant soft x-ray microscopy, which we have previously reported in Refs. 17 and 18. Furthermore, numerical model suggest that the domain wall is Bloch-like in the center of the films but broaden and transitions to more Néel-like towards the surface forming closure domains. The modeling shows that while the Bloch-like centers have an equal population of the two helicities, the Néel-like part of the walls will have the same helicity at the top of the film and the opposite helicity at the bottom of the film.

Utilizing a broad-band coplanar waveguide technique, we have studied the resonant behavior of the magnetic textures as a function of magnetic field, temperature and film thickness. In addition, we have performed micromagnetic simulations under time-dependent *r.f.* magnetic fields to compare to the experimental results and determine the resonant modes and spin configurations of these magnetic domains. Our modeling results in frequency-field dispersions that agree with our experimental observations, and suggest the existence of new localized spin wave modes. Unlike DMI skyrmions, we predict dipole skyrmions exhibit non-uniform dynamics across the film thickness that are dependent on the helicity of the chiral domain-wall. Furthermore, we show the resonant spectra is not sensitive to the ordering of the dipole skyrmions into arrays.

## II.    Methods

The Fe/Gd multilayer films were grown at ambient temperature by d.c. magnetron sputtering in an Ar pressure of 3mTorr. To protect the films against corrosion a 5-nm-thick Ta seed and capping layer was used. Films are deposited on 50-nm and 200-nm-thick $Si_3N_4$ membranes for Lorentz TEM, soft X-ray scattering and X-ray microscopy measurements, as well as, native oxide coated Si substrate for magnetic characterization. The temperature dependent magnetic loops are measured using a Quantum Design Versalab and Physical Property Measurement System (PPMS) with a vibrating sample magnetometer, from which the saturation magnetizations and saturation fields are obtained.

The magnetic excitations of the Fe/Gd films are extracted by ferromagnetic resonance (FMR) measurements performed using a custom-built probe that is inserted into a Quantum Design PPMS cryostat for temperatures from 2 K to 400 K and d.c. magnetic field $H_{dc}$ up to 9 T. We use the fixed frequency FMR technique [26, 27] to study the resonant properties of the films, where a broad-band coplanar waveguide is affixed to the bottom of the custom probe (Supp. Fig. S1 [28]). In this geometry (shown schematically in Fig. 1b), the *r.f.* magnetic field ($h_{rf}$) pulse lies on the plane of the waveguide and the magnetic specimen is affixed on top of the waveguide. The absorption spectra are captured using an Agilent Vector Network Analyzer E8363B at fixed frequencies while the applied magnetic field ($H_{dc}$) is scanned.

## III.  Results

### A.  Temperature dependence of resonant modes.

We initially investigated the resonant properties of a [Fe(0.36nm)/Gd(0.40nm)]x80 multilayer [17,18]. This sample exhibits disordered stripe domains below magnetic saturation from room temperature down to 180K. From 180K to 100K the sample shows a transition from disordered stripes to a dipole skyrmion lattice when a magnetic field is applied perpendicular to the film [17,18]. Measuring the resonant properties in these two temperature ranges allows us to distinguish the resonant properties that arise from a dipole skyrmion phase from that of the stripe phase. When we probe the system with a fixed pump field $h_{rf}$ parallel to the film while sweeping a perpendicular $H_{dc}$ we obtain absorption spectra $\Delta|S_{12}|$ that shows up to four resonance peaks at each fixed frequency.

Figure 1c shows the field dependence of $\Delta|S_{12}|$ for 14 different frequencies measured at 160K while sweeping $H_{dc}$ from negative to positive magnetic saturation (In Fig. 1c we only show the absorption spectra from zero-field to magnetic saturation). At each frequency, the absorption peak occurring at the highest resonate field is the Kittel resonance peak [29, 30] arising from the precession of the homogeneous magnetization state of the film when the magnetic film is saturated. The absorption peaks observed at lower resonant fields are associated with non-homogeneous excitations and/or non-homogeneous magnetic states, *i.e.* stripe domains, bubbles and dipole skyrmions. By fitting the Kittel peak with a Lorentzian profile we obtain the field dependence of the precession frequency (Fig. 2a). The resonant frequency varies linearly with applied field as expected from the Kittel formula [29, 30], $f = \frac{\gamma_e}{2\pi}(H_{dc} - 4\pi M_{eff})$, where $4\pi M_{eff} = 4\pi M_S - H_k$ and $M_S$ is the saturation magnetization and $H_k$ is the perpendicular magnetic anisotropy field. This linear behavior shifts to higher resonant fields as the temperature is decreased (Fig. 2a). Extrapolating the Kittel peak to $f = 0$ GHz gives the effective magnetization $4\pi M_{eff}$ (Fig. 2b). From the fact that $4\pi M_{eff}$ is positive we know the film possesses a weak perpendicular magnetic anisotropy which is less than the thin-film shape anisotropy and $4\pi M_{eff}$ becomes more positive as we decrease $T$. From $M_S$ (Supp. Fig. S2 [28]) and $4\pi M_{eff}$ (Fig. 2b), we determine the effective perpendicular magnetic anisotropy $K_{eff}$ (which includes both the intrinsic perpendicular magnetic anisotropy $K_U$ and the thin-film shape anisotropy $-2\pi M_S^2$) as plotted in Fig. 2c decreases as we reduce the temperature (Fig. 2c).

The Kittel resonance peaks line-width is typically small ($\Delta H \sim 100$ Oe) which suggests the effective Gilbert damping $\alpha_{eff}$ is low. By fitting the linewidth, $\Delta H$, as a function of the FMR frequency to $\Delta H = \frac{4\pi f \cdot \alpha_{eff}}{\gamma_e} + \Delta H_o$, where $\Delta H_o$ is the inhomogeneous broadening and $\gamma_e$ is the $e^-$ gyromagnetic ratio, we find that the effective Gilbert damping $\alpha_{eff}$ decreases monotonically from 0.0167 to 0.0078 with temperature (Fig. 2b). These values are in good agreement with reported values for Gd-rich Fe-Gd ferrimagnets [31, 32].

From a complete data set similar to that shown in Fig. 1c measured from 240K to 80K, we have extracted the resonant modes from the FMR spectra by fitting the peaks with Lorentzian profiles and plotted their normalized peak intensity in frequency-field (*f-H*) phase diagrams in Fig. 3. The field history was the same as that described for Fig. 1d, sweeping the field from negative saturation to positive saturation while plotting only the resonant values from remanence to positive saturation. We have also investigated the field effects on the resonant spectra (Supp. Fig. S3 [28]), as a function of temperature, which is described in Supplementary Section 3 [28]. For each resonance spectra, the intensity of each peak (normalized to the highest intensity peak at each fixed frequency) is given by the size of the symbol. We see that at each temperature there are two resonant mode regions. The first is above magnetic saturation and is described by the Kittel formula (as discussed above). The second region is below the saturation field where there are typically 2 or 3 mode branches that increase in frequency with decreasing field. In most cases, above *f* = 2.5 GHz the highest absorption peak intensity is given by the uniform mode. However, at lower frequencies we observe very high-intensity modes resulting from the low-field magnetic textures.

The resonant spectra resulting from the non-homogeneous magnetic textures shows that the domain morphology evolves as we reduce the temperature. From 240K to 180K, the multiple resonant modes appearing below *f* ~ 2.5 GHz become a single mode branch below 200K. Further a mode branch appearing at *f* ~ 4.5 GHz becomes more prominent as the temperature is reduced. From 160K to 80K, the two mode branches below magnetic saturation are similar and relatively insensitive to temperature. Using magnetic phase maps constructed from real-space imaging and reciprocal-space scattering techniques described in Ref. [18], we can overlay on the *f-H* diagrams the specific magnetic textures at each field and temperature. We have identified up to five different field regions corresponding to distinct magnetic structures: (I) disordered stripes, (II) coexisting stripes and dipole skyrmions, (III) skyrmion lattice, (IV) disordered isolated skyrmions and (V) magnetic saturation and these magnetic states are shown on Fig. 3. These regions are delimited by dashed-lines which corresponds to the magnetic fields at which distinct magnetic structures exist. At temperatures where no skyrmion phase is observed, 240K to 200K (Fig. 3a-c), we observe the low-field mode branches decrease in frequency with increasing field and reach a minimum that corresponds to the saturation field which can be estimated from extrapolating the linear Kittel peak in the high-field region. As we decrease the temperature we see the extrapolation of the low-field modes to zero frequency occurs at lower magnetic fields than the saturation field, forming an effective gap in magnetic field between the low- and high-field behavior. As the temperature is further reduced this gap increases. Here, the magnetic field at which the mode branches reach zero frequency does not change considerably with temperature and instead the gap increase is associated with the Kittel peak shifting to higher fields as the magnetization increases and the anisotropy decreases. Also we observe that there is no systematic splitting of the low-field mode-branches at magnetic fields where the skyrmion phase is present in contrast to what is seen in materials where Bloch-type skyrmions are stabilized by DMI [20,21]. Since dipolar skyrmions in Fe/Gd films are not a result of a field-driven phase transition from a conical to skyrmion phase, like in DMI materials, one might expect a continuous variation of the modes from the stripe to skyrmion phases.

At low temperatures, 120K to 80K, we also observe a secondary mode-branch appears above magnetic saturation (Fig. 3g-i). This mode occurs at lower resonant fields than the Kittel peak and has much lower absorption intensity. The spacing in resonant field between the second mode branch and the Kittel mode branch is nearly uniform in frequency. Usually, a single resonance is expected above magnetic saturation, $e.g.$ the Kittel resonance, yet variations in surface anisotropy in the thin film or non-uniform $r.f.$ field amplitudes in the film can result in the excitation of a standing spin-wave resonance [33]. Therefore, the Kittel mode branch corresponds to the finite spin wave resonance with wave number $k = 0$ and the second mode branch is the $k = 1$ spin wave resonance. The difference in resonant field between these two spin waves [34-36] is given by

$$4\pi M_{eff} - H_o = \frac{2A_{ex}}{M_S}k^2, \quad k = \frac{p\pi}{t_m} \qquad (1)$$

where $4\pi M_{eff}$ is the effective magnetization arising from the $k = 0$ resonant mode, $H_o$ is the effective magnetization of the $k = 1$ spin wave resonance, $A_{ex}$ is the exchange stiffness, $p$ is the integer number of nodes, and $t_m$ is the film thickness. We find the exchange stiffness varies from $A_{ex}$= 5.0x10$^{-7}$ to 6.7x10$^{-7}$ erg/cm, depending on the temperature, which is consistent with numerical values predicted in Ref. 18.

## B. Thickness dependence of the resonant modes.

We studied the dependence of the resonant modes on the Fe/Gd multilayers thickness for [Fe(0.34nm)/Gd(0.41nm)]xN with different number of bilayers N = 80 and 120. Using full-field transmission soft X-ray microscopy, at the Fe $L_3$ (706 eV) absorption edge, we first investigated the field-dependent domain morphology at room temperature. The latter two structures (N=80 and 120) are the only ones that show perpendicular magnetic domains, where [Fe(0.34nm)/Gd(0.41nm)]x80 depicts stripe domains that pinch into cylindrical domains (Fig. 4a) and [Fe(0.34nm)/Gd(0.41nm)]x120 shows disordered stripe domains that collapse into skyrmions which arrange into a close packing lattice (Fig. 4b). We note here that the increase of the Gd thickness to 0.41 nm compared to the 0.40 nm (Figs. 1-3) shifts the temperature range where skyrmions are observed to room temperature [18].

When the observed domains states are probed with an in-plane *r.f.* field $h_{rf}$, we find the resulting *f-H* dispersions are qualitatively similar even though the domain morphologies are distinct (Figs. 4c and d). Above magnetic saturation, the comparable effective magnetization $4\pi M_{eff}$ suggests the films have the same average magnetic properties (Figs. 4c and d inserts); hence, variations in domain morphology and resonant modes are a result of magnetostatic energy differences arising from different total film thicknesses. Comparing the *f-H* dispersions of [Fe(0.34nm)/Gd(0.40nm)]x80 and [Fe(0.34nm)/Gd(0.41nm)]x80, when a skyrmion phase is present (Figs. 3d-h and 4c), we find that both display almost identical resonant mode braches above and below magnetic saturation. The ordering of skyrmions, whether in a close-packing lattice (Fig. 4b) or skyrmions arranged in a line (Fig. 4a), is not reflected in the *f-H* dispersions. As the number of bilayer repetitions is increased, [Fe(0.34nm) /Gd(0.41nm)]x120, we observe the resonant mode branches shift closer to magnetic saturation and collapse near the saturation field (Fig. 4d). At and above magnetic saturation (Figs. 4c and d), the resonance frequency of [Fe(0.34nm)/Gd(0.41nm)]x80 and [Fe(0.34nm)/Gd(0.41nm)]x120 linearly increase with field as expected for the uniform magnetization.

## C. Comparison to micromagnetic modeling

### 1. Field dependent domains states and their resonant modes

To determine the micromagnetic nature of the modes observed, we performed numerical simulations of the Landau-Lifshitz-Gilbert (LLG) equation under a combination of *d.c.* fields and time-dependent *a.c.* magnetic fields, utilizing the FASTMag solver [37]. We use magnetic parameters where a skyrmion lattice forms numerically (from Ref. [18]: $M_S = 400$ emu/cm$^3$, $K_U = 4 \times 10^5$ erg/cm$^3$ and $A = 5 \times 10^{-7}$ erg/cm) which are consistent with the magnetic properties of the [Fe(0.36nm)/Gd(0.40nm)]x80 sample. First, we explored the field-dependent equilibrium states that form in a slab with volume of 2μm x 2μm x 80nm that is discretized with 10-nm tetrahedra. The simulation is first saturated with a perpendicular magnetic field and then the field is reduced to $H_z = 0$ Oe and allowed to reach equilibrium. Then a perpendicular *d.c.* magnetic field is increased from $H_z = 0$ Oe to 4000 Oe in discrete steps and the slab is allowed to reach an equilibrium state in 30 ns. The field history is similar to the experimental protocol used for the experimental results in Supplementary Section 3 [28]. After reducing to zero field we find the remanent state consists of stripe domains with random in-plane order (Fig. 5a, $m_x$ at z = 0nm) that is similar to experimental observations [18]. As discussed in Ref. 18 inspecting a cross-section of the slab, across the thickness, reveals a domain morphology consisting of perpendicular domains with domain walls that are Bloch-like in the center of film and become more Néel-like towards each surface forming closure domains, *e.g.* Néel caps. Given the transmission geometry nature of the experimental imaging techniques used and the symmetry of the Néel caps they images are primarily sensitive to the Bloch-like, full experimental verification of the domain wall structure will require a 3-dimensional imaging of domain walls [38], whereas sole confirmation of the Néel caps could be performed utilizing surface sensitive microscopy techniques [39-41]. As a perpendicular field is applied, the stripe domains begin to collapse into an equal population of skyrmions with two possible helicities S = 1, $\gamma = \pm\pi/2$ (Figs. 5a, 8, Supp. Fig. S4 [28]). Increasing the field further, the skyrmions arrange into a close packing lattice that spans from $H_z = 1600$ Oe to 2400 Oe (Fig. 5). After $H_z = 2400$ Oe, the skyrmions become disordered and begin to collapse as the field is increased to magnetic saturation. Overall the field-dependent domain morphologies are in good agreement with our experimental observations in Fig. 4 and Ref. 18.

To theoretically determine the dynamic response, we first perturbed the magnetization states with a pulsed field along the *x*-axis or *z*-axis, with an amplitude of $h_x = h_z = 100$ Oe, and monitor how the system returns to equilibrium (Fig 5b). To obtain the susceptibility $\chi(\omega)$, the dynamics of the magnetization in time-domain are recorded when exposed to the pulsed field. Afterwards, both the excitation field and the magnetization response are Fourier transformed using a fast Fourier transform (FFT) approach. It follows that the susceptibility in frequency domain is given by:

$$\chi(\omega) = \frac{M(\omega)}{H(\omega)} = \chi'(\omega) + i\chi''(\omega) \tag{2}$$

where $\omega$ is the frequency, $M(\omega)$ magnetization and $H(\omega)$ field pulse in frequency domain; furthermore, $\chi'(\omega)$ is the real part of $\chi(\omega)$ that reflects the sensitivity of the magnetization under a field and $\chi''(\omega)$ is the imaginary part of $\chi(\omega)$ that reflects the dissipation of energy being absorbed. By fitting the various $\chi''(\omega)$ resonant peaks with Lorentzian profiles from both perturbation geometries, we obtain a *f-H* dispersion with up to three mode branches appearing in the non-homogeneous magnetization region that are similar in frequency to the lower frequency mode branch observed experimentally and two mode branches in the saturated state (Fig. 5c) where one is the Kittel mode and the less intense mode is an edge mode arising from the finite modelled area (Supp. Fig. S5 [28]) and is not related to the $k = 1$ mode described earlier. We have also explored the sensitivity of the skyrmion resonant modes to variations in exchange parameter (Supp. Fig. S6 [28]), which shows that the intensity of the resonant peaks of the lower (*f* ~1.3 GHz) and upper (*f* ~ 2.6 GHz) mode branches varies significantly with exchange parameter. Finally, in Supplementary Section 7 [28] we investigated the local susceptibility spectra (Supp. Fig. S7 [28]) resulting from magnetic features that appear in a skyrmion closed packed lattice.

## 2. Localized spin wave modes of dipole skyrmions

To resolve specific dynamic magnetization configurations that appear in the various mode-branches in the susceptibility spectra we apply a sinusoidal *a.c.* magnetic field at a given resonant frequency and record snapshots of the magnetic domain evolution during a sinusoidal field cycle [22] (Fig. 6a). Our modeling suggests a variety of localized spin wave modes exist and depends on the helicity of the dipole skyrmions (Fig. 6a).

Here we describe the spin-wave configurations resulting from an in-plane sinusoidal field $h_x^{rf}$ at resonant frequencies: *f* ~ 1.3 GHz, 1.8 GHz, 2.6 GHz at $H_z$ = 2000 Oe (Figs. 5c and 6a) which depicts the dynamics of a skyrmion lattice that forms at this field. At each resonant frequency the magnetic features resonate with the following properties: (i) the localized spin-wave is asymmetrical on the top and bottom of the skyrmion domain and (ii) a skyrmion with opposite helicity shows the inverse spin-wave mode that is degenerate in frequency. To show this complex behavior, we detail the displacement of the various magnetic textures using contour lines at different quadrant positions for a single sinusoidal pulse (Fig. 6b). Figure 6a shows the spin-wave motion of the out-of-plane magnetization at three different depth positions of the slab (top surface, center and bottom surface) for both helicity skyrmions (S = 1, γ = ±π/2). The oscillation of the Néel caps and Bloch-line under a sinusoidal field $h_{x,z}^{rf}$ is detailed in the Supplementary Movies 1-4 [42]. Given the symmetry of the spin-wave mode response for skyrmions with opposite helicity, we describe in detail only the behavior for an (S = 1, γ = -π/2) skyrmion. First, the localized spin-wave associated with the $f \sim 1.3$ GHz branch consists of a linear displacement of the skyrmion top from right-to-left-to-right position during a sinusoidal pulse cycle. Given the back and forth displacement during the pulse field cycle, we refer to it as localized 'linear' spin-wave mode. The skyrmion bottom exhibits a linear displacement from right-to-left-to-right that is rotated by ~135°. In the case of the $f \sim 1.8$ GHz spin-wave, a similar linear displacement, from right-to-left-to-right, is observed on the top of the skyrmion, however, the bottom part of the skyrmion rotates in clock-wise motion. Lastly, the $f \sim 2.6$GHz spin-wave displacement is similar to that of the $f \sim 1.8$ GHz mode with the exception that the linear displacement occurs at a small tilt angle away from the perturbation field along the x-axis. From inspecting the spin-wave motion for all three mode-branches, we find that greater displacement is towards the surface of the film which is the region where the Néel caps have a larger presence.

The spin-wave mode associated with a perturbation along the z-axis, consists of a breathing-like displacement where the skyrmion expands and contracts under the sinusoidal field. This spin-wave mode is similar to that reported by Mochizuki, *et al.* [Ref. 22] in DMI skyrmion materials. Here, the skyrmion domain expands/contracts non-uniformly throughout the thickness of the film: for an (S = 1, γ = -π/2) skyrmion, the core contracts and expands at the top of the slab; whereas at the bottom, the core expands and then contracts. The inverse dynamics is observed for a dipole skyrmion with opposite helicity. Altogether, the breathing displacement for a skyrmion is weaker in amplitude compared to the motion resulting from perturbing the same equilibrium state with a sinusoidal field $h_x^{rf}$ pulse. In the case of the Néel caps, we also observe a breathing-like displacement that also twists from right-to-left around the skyrmion. This behavior is most noticeable around the center of the film (Supp. Movie 4 [42]).

### 3. Field effects on domain states and their resonant modes

The numerical modelling accurately replicates the lower frequency mode-branches from the *f-H* dispersions detailed for the [Fe(0.36nm)/Gd(0.4nm)]x80 multilayer, where a skyrmion phase is present although we do not observe three distinct modes. It does not, however, reproduce the experimental mode branch occurring at $f \sim 4.4$ GHz. For this reason, we also considered modeling the response of a domain morphology consisting of a combination of bubbles and skyrmions because it is possible to experimentally stabilize both textures in amorphous Fe/Gd multilayers [17, 18]. We achieve this by applying a constant small in-plane field along the x-direction with an amplitude of $H_x = 60$ Oe while a perpendicular magnetic field is varied. We find that disordered stripe domains become ordered and align transverse to the $H_x$ field at $H_z = 0$ Oe (Fig. 7a). Examining the in-plane magnetization at the top and the center of the slab reveals the Néel caps align in the direction of the constant $H_x$ field, which results in the Bloch-line pointing orthogonal to the fixed $H_x$ in-plane field (Fig. 7a, Supp. Fig. S8 [28]). The ordering of the Néel caps and Bloch-line under an in-plane field is described further in the Supplementary Section 8 [28].

Subsequently, as the perpendicular field is applied while keeping the constant in-plane field, the stripe domains begin to collapse into cylindrical-like domains (Fig. 7a). Unlike in the results shown in Fig. 5a where a majority of the magnetic features are skyrmions we observe that the application of an in-plane field results primarily in the formation of bubble domains (S = 0) and some skyrmions which all arrange into a close-packing lattice that exists from $H_z = 1600$ Oe to $H_z = 2400$ Oe. We observe the Bloch-line for the bubbles aligns in the direction of the in-plane $H_x$ field as opposed to circling the domain as seen in the skyrmion phase. As the perpendicular field is increased further, the bubbles and skyrmions become disordered and these textures begin to dissipate toward magnetic saturation with bubbles collapsing at lower fields than skyrmions (Fig. 7a). Given the topological nature of skyrmions, a higher annihilation field is required compared to bubbles. Experimentally we also observed that with in-plane fields can also result in bound skyrmion pairs of opposite chirality [17].

Applying the same *a.c.* field perturbations, as described before, results in a comparable susceptibility spectrum like the one elicited from equilibrium states without a constant in-plane field (Fig. 5c) together with an additional mode branch appearing at $f \sim 4.4$ GHz (Fig. 7b). In the *f-H* diagram, we overlay the different magnetic textures that form under a magnetic field based on results like those in Fig. 7a. The magnetization states include: (I) stripes, (II) coexisting stripes and cylindrical domains, (III) ordered lattice of cylindrical domains, (IV) disordered cylindrical domains and a (V) Kittel region. We find the mode-branch at $f \sim 4.4$ GHz is fairly constant with increasing field even as the magnetic textures change from stripes to an ordered lattice of cylindrical-like domains and decays sharply in resonant frequency and intensity when the slab exhibits disordered cylindrical-like domains.

When exposing the magnetic configurations that form at $H_z = 2000$ Oe to a sinusoidal a.c. field along either $h_x^{rf}$ or $h_z^{rf}$ at one of the resonant frequencies ($f \sim 1.4$ GHz, 1.9 GHz, 2.4 GHz and 4.4 GHz) we can determine the spin-wave configurations that appear for both chirality cylindrical domains. The skyrmions (S = 1, $\gamma = \pm \pi/2$) exhibit the equivalent localized spin-wave modes previously described at similar resonant frequency. In contrast the bubble domains (S = 0) exhibit the same localized spin wave dynamics at all of the four resonant frequencies. The bubble domains exhibit a linear displacement of the core in the direction of the oscillating $h_x^{rf}$ field (Fig. 7c). The top of the bubble shifts from left-to-right-to-left, whereas at the bottom, the motion is from right-to-left-to-right. Considering the alignment of a magnetic bubble's Néel caps and Bloch-line in the direction of the in-plane $H_x$ field, the linear displacement of the bubble domain is anticipated. Probing the textures with a perpendicular sinusoidal field $h_z^{rf}$ results in a similar breathing-like spin wave excitation for both bubbles and skyrmions, as detailed before.

The experimental observation of the mode-branch at $f \sim 4.4$ GHz suggests the formation of bubble domains in the Fe/Gd films which could be attributed to different mechanisms, such as: (*i*) the in-plane field produced by the coplanar waveguide is sufficient to orient the Bloch-line of the stripe domains as the field is reduced from magnetic saturation, (*ii*) alternatively a tilt of the sample with respect to the coplanar waveguide could similarly result in the application of an additional in-plane field to the film. All of these scenarios are possible given the Fe/Gd film has soft magnetic properties (Fig. 2). As a result, the *f-H* dispersions detailed in Figs. 3, 4 and Supp. Fig. S3 [28] are predicted to detail the resonance behavior of a mixture of magnetic bubbles and dipole skyrmions.

## IV. **Discussion**

Dipole skyrmions (Fig. 8a, b) and achiral bubbles (Fig. 8c) exhibit similar domain structure (*e.g.* perpendicular domain with asymmetric Néel caps at the top and bottom of the film and a Bloch-line center) with the exemption of a domain wall with continuous vorticity. For both topological magnetic features, the Néel caps at the top point inwards to the cylindrical feature and outwards at the bottom of the magnetic feature (Fig. 8a-c) for positive applied fields. At the center of the film, a dipole skyrmion possesses a Bloch-line that continually wraps around the magnetic feature (Fig. 8a, b) whereas a bubble domain has two Bloch-point discontinuities at which the magnetic spins circulating the cylindrical texture change vorticity (Fig. 8c). Furthermore, Néel caps of dipole skyrmions possess chirality analogous to the Bloch-line at the center of the magnetic feature. In the case of an ($S = 1$, $\gamma = -\pi/2$) skyrmion the Néel caps have a slight clockwise component (Fig. 8a), whereas an ($S = 1$, $\gamma = +\pi/2$) skyrmion possesses Néel caps with a slight counter-clockwise component (Fig, 8b). On the other hand, bubble domains have Néel caps that are purely radial close to the Bloch-point discontinuity and exhibits a vorticity component away from the Bloch-points that changes vorticity after passing a Bloch-point (Fig. 8c). We will argue that the uniform chirality throughout a dipole skyrmion is the source of the specific localized dynamics.

The observation of non-uniform dynamics across the thickness of dipole skyrmions and bubbles domains we believe is the result of asymmetric Néel caps that extend from the center of the film to the top and bottom surfaces. To illustrate this, we consider a purely radial achiral bubble ($S = 0$) whose simplified magnetic structure is detailed at different depths in Fig. 9. At the top and bottom surface, the Néel cap magnetic spins are parallel and orthogonal to the *r.f.* field. When the achiral bubble is perturbed with an in-plane sinusoidal field: (i) the magnetic spins along the direction of the *r.f.* field exhibit negligible field torque $\mu \times h_{rf}$ and will only result in an increase/decrease in magnitude relevant to the maxima/minima of the sinusoidal field (Fig. 9). (ii) Comparable interpretations can be reached for the Bloch-line center of the film in which the magnetic spins are also aligned in the direction of the *r.f.* field. (iii) The orthogonal magnetic spins will exhibit field torques that displaces the surface magnetic spins inward/outward the magnetic texture (Fig. 9). As a result of the latter dynamic behavior, the magnetic spins will exhibit two more competing interactions: at the surface the stray field prefers to align the magnetic spins into the bubble domain to reduce the magnetostatic energy, whereas the exchange interaction prefers the magnetic spins be aligned parallel to each other. Overall, these competing interactions will drive the orthogonal magnetic spins in the Néel caps, relative to the *r.f.* field, into precession. At the sinusoidal field nulls, $h_{rf} = 0$, the orthogonal magnetic spins will be oriented in the same direction which results in the Néel caps displacing uniformly at each surface. Since inverse field torques are observed at opposite Néel caps, the displacement is asymmetric across the thickness of the bubble.

The perpendicular magnetic spins across thickness of the purely radial achiral bubble will also exhibit a field torque that causes the spins to shift orthogonally (Fig. 9). The precession of these magnetic spins results from the demagnetization field that wants to realign the canted magnetic spins downward to reduce the magnetostatic energy. Since the domain wall moves in the direction of the in-plane sinusoidal field, the precessional motion of perpendicular magnetic domains are confined to the oblique path set by the domain wall. Moreover, the observation of asymmetric dynamics from perpendicular magnetic spins is rooted to the opposite displacement of Néel caps that exist away from the center of the magnetic bubble.

Given the Néel and Bloch nature of dipole skyrmions, the predicted dynamics arising from perturbing the chiral magnetic textures with a microwave field has characteristics associated with both Néel-type and Bloch-type DMI skyrmions. The breathing mode observed in dipole skyrmions exhibits a skyrmion core expansion and contraction similar to Bloch-type DMI skyrmions [22] and the magnetic spins orthogonal to the *r.f.* field fluctuate as Néel-type DMI skyrmions [25]. Under in-plane microwave fields, the dipole skyrmions exhibit displacements that are dependent on the resonant frequency, like DMI skyrmions, but also on the helicity of the domain wall. A dipole skyrmion possesses a vortex-like magnetic spin configuration at the surfaces which is comparable to Bloch-type DMI skyrmions [22], reason for which the surface localized spin wave dynamics between both topological textures are comparable.

In experiments, we predict Bloch-type DMI skyrmions will show localized spin wave modes closer to those resulting from dipole skyrmions at the surface of the magnet. Current numerical models of DMI skyrmion dynamics assume a two-dimensional lattice structure [22], but Bloch-type DMI skyrmions appear in thick films or bulk materials where the magnetostatic energy plays a strong role in the formation of magnetic domains and can modify the dynamics of these textures especially at the surface due to the presence of strong stray fields.

Differences in microwave field dynamics between DMI and dipole skyrmions arise from the arrangement of magnetic spins across the thickness. A DMI skyrmion only varies radially outward in a Néel-way or Bloch-way, and because of this we expect homogenous microwave field dynamics across the skyrmion depth. However, the magnetic spin configuration of dipole skyrmions varies notably with thickness, as a result, different localized spin wave displacements are observed across throughout the chiral texture. Furthermore, Mochizuki, *et al*. [22] predicts that the spin wave dynamics of Bloch-type DMI skyrmions is independent on their winding number or the sign of the DMI constant, *i.e.* clockwise and counter-clockwise Bloch-type DMI skyrmions will show the same microwave field excitations at the skyrmion resonance frequency; whereas, dipole skyrmions with opposite helicity show the inverse dynamics across the thickness. Finally, when comparing the dynamics from magnetic bubbles and dipole skyrmions it is evident that the chiral nature of the dipole skyrmions is responsible for the complex excitations under microwave fields.

## V.    Conclusion

We have experimentally measured and modeled the resonance properties of dipole skyrmions and bubble domains that form in amorphous Fe/Gd films through temperature and thickness dependent studies. First, the continuous decay of the mode branches associated with dipole skyrmions shows that these textures cannot be identified directly in *f-H* dispersions given there is no splitting of the mode branches at magnetic fields where the skyrmion phase forms out of the stripe phase. Similar continuous mode branches have been observed for Néel-type DMI skyrmions [25]. Experimental results also show that the arrangement of dipole skyrmions, whether in closed-packed lattices or isolated skyrmions, is not reflected in the resonant spectra.

Through simulations, we predict the magnetization dynamics resulting from bubble domains and dipole skyrmions are distinct when these textures are probed with microwave fields. In the case of dipole skyrmions, we have identified four distinct localized spin wave modes when these textures are probe with either in-plane or perpendicular microwave fields. Complex dynamics are expected across the thickness of the dipole skyrmion where opposite helicity textures show inverse dynamics at any resonant frequency. These localized resonant modes are comparable to those predicted in DMI skyrmions. In overall, it appears topology of a skyrmion is responsible for the non-trivial dynamics observed in dipole and DMI skyrmions and there is no contribution from the physical stabilization mechanism. In future technologies, the different dynamics on the surface of a dipole skyrmion could be exploited as an additional degree of readability to distinguish the two possible helicities, thus enabling each helicity skyrmion a binary bit.

**Acknowledgements.**

Work at University of California–San Diego including materials synthesis and characterization, participation in synchrotron measurements and modeling was supported by U.S. Department of Energy (DOE), Office of Basic Energy Sciences (BES) under Award No. DE-SC0003678). Work at University of Oregon was supported by the DOE, Office of Science, BES under Award No. DE-SC0010466. Work at the Advanced Light Source, Lawrence Berkeley National Lab (LBNL) was supported by the Director, Office of Science, BES, of the DOE (Contract No. DE-AC02- 05CH11231). S.A.M. acknowledges the support from the Department of Defense (DoD) through the Science, Mathematics & Research for Transformation (SMART) Scholarship. B.J.M. and J.J.C. gratefully acknowledge the use of CAMCOR facilities, which have been purchased with a combination of federal and state funding. M.-Y. I. acknowledges support by Leading Foreign Research Institute Recruitment Program through the National Research Foundation (NRF) of Korea funded by the Ministry of Education, Science and Technology (MEST) (Grants No. 2012K1A4A3053565 and No. 2014R1A2A2A01003709). S.D.K. and P.F. acknowledge support by the Director, Office of Science, BES, Materials Sciences and Engineering Division, of the DOE under Contract No. DE-AC02-05-CH11231 within the Nonequilibrium Magnetic Materials Program (No. KC2204) at LBNL.

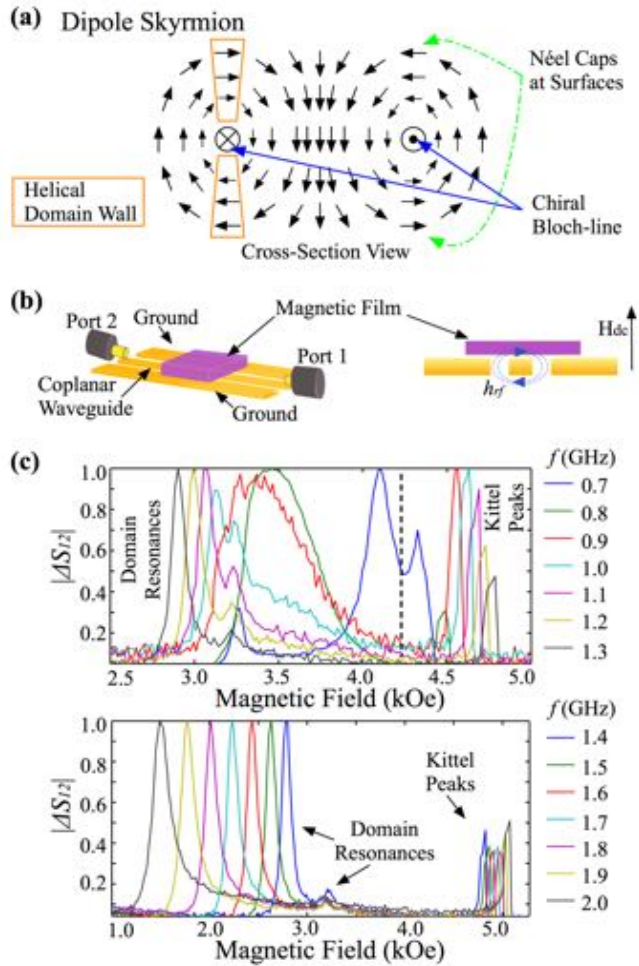

**Figure 1. (a)** Schematic shows the cross-section view of a dipole skyrmion where the domain wall has a helical structure across the skyrmion depth with opposite Néel caps at the top and bottom surface and a chiral Bloch-line center. **(b)** Schematic of the experimental set-up with the film on a coplanar waveguide. **(c)** The normalized absorption spectra, measured at 160K, for a [Fe (0.36nm) / Gd (0.4nm)]x80 multilayer. The spectra are obtained while sweeping a perpendicular magnetic field, from $H_z$= -10 kOe to +10 kOe, at a fixed frequency $h_{rf}$ pump field. Resonances at/above the saturation field exhibit a smaller line-width than resonances elicited by non-homogenous magnetic textures. The dashed line in **(c)** indicates the magnetic saturation field: resonances appearing above this field known as Kittel peaks, and resonant peaks appearing below magnetic saturation are denoted domain resonances.

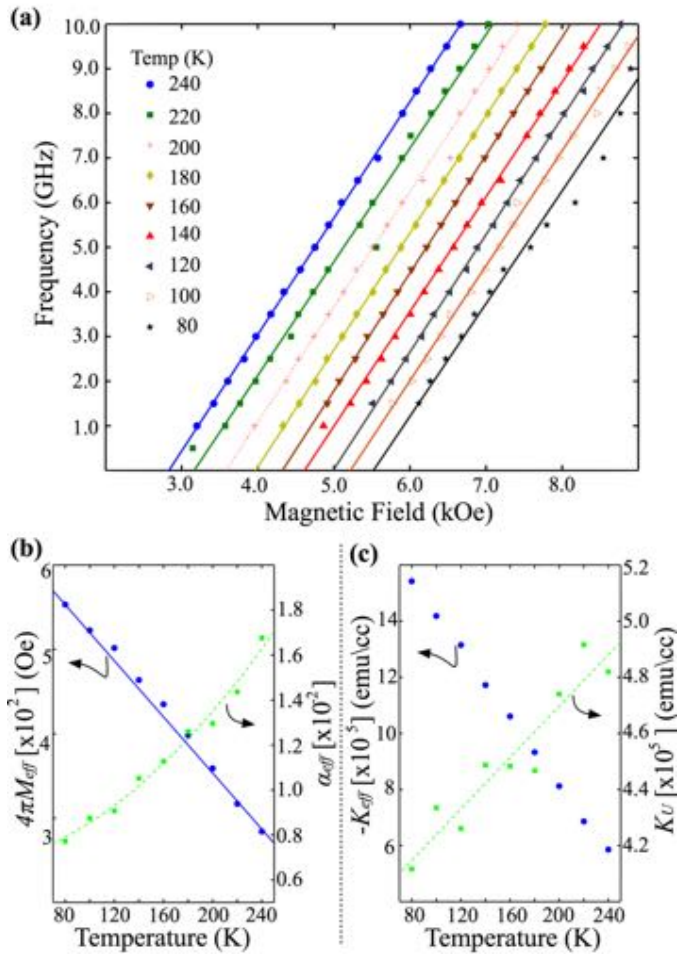

**Figure 2.** Resonance and magnetic properties of a [Fe (0.36nm) / Gd (0.4nm)]x80 multilayer **(a)** The field dependence of the resonance resulting from the precession of the homogenous magnetization, e.g. Kittel mode, is shown from 240K to 80K. **(b)** Fitting results from **(a)** and the line-width of the Kittel peaks, like those in Figure 1c, we determine the temperature dependence of the effective magnetization and effective damping. **(c)** The extracted effective and intrinsic uniaxial anisotropy calculated from results in **(b)** and Supplementary Figure S2 [28].

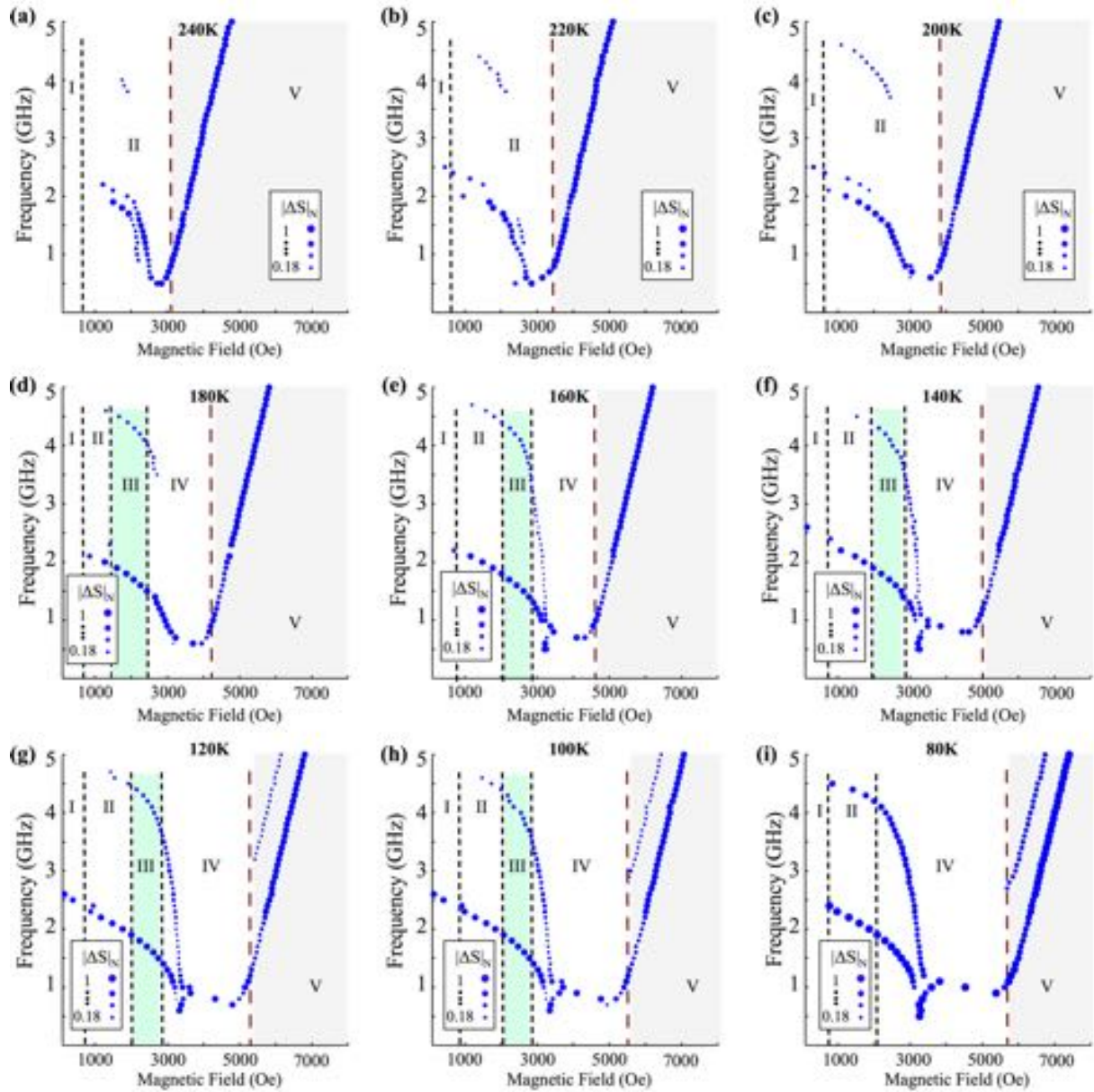

**Figure 3.** Temperature dependence of the resonant modes of a [Fe (0.36nm) / Gd (0.40nm)]x80 multilayer. The frequency-field *f-H* dispersions illustrate the resonant modes that exist at temperatures ranging from 240K down to 80K. These *f-H* dispersions were constructed from measurements were the perpendicular magnetic field is swept from $H_z$= -10 kOe to +10 kOe at each fixed frequency, and detail the resonances observed from zero-field to 8000 Oe. At each temperature, several regions are identified that correspond to specific domain morphologies: (I) stripes, (II) coexisting stripes and skyrmions, (III) skyrmion lattice, (IV) disordered skyrmions, and (V) Kittel region as described in Ref. 18. These regions are delimited with dashed lines in the *f-H* dispersions. The saturation field $H_{sat}$ is obtained from magnetic hysteresis measurements performed in the same geometry.

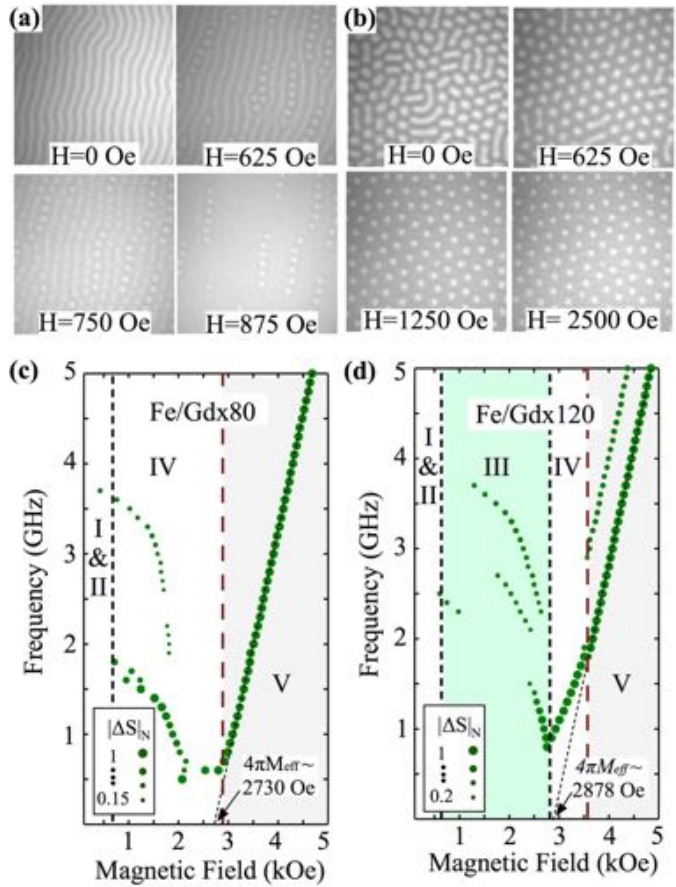

**Figure 4.** Film thickness dependence of the resonant modes at room temperature. **(a, b)** Images of the domain morphology field evolution of a [Fe (0.34nm) /Gd(0.41nm)]x80 and [Fe (0.34nm) /Gd(0.41nm)]x120 under positive perpendicular magnetic fields. The images are obtained with transmission X-ray microscopy along the Fe $L_3$ absorption edge (708 eV), where white features depict magnetization (-$m_z$) and gray regions exhibit magnetization (+$m_z$). **(c, d)** The $f$-$H$ dispersions for the latter Fe/Gd films show the field-varying resonant modes observed at room temperature. From images like **(a, b)** we correlate the modes to specific domain morphologies: (I) stripes, (II) coexisting stripes and skyrmions, (III) skyrmion lattice, (IV) disordered skyrmions and (V) Kittel region. These regions are delimited with dashed lines in the $f$-$H$ dispersions.

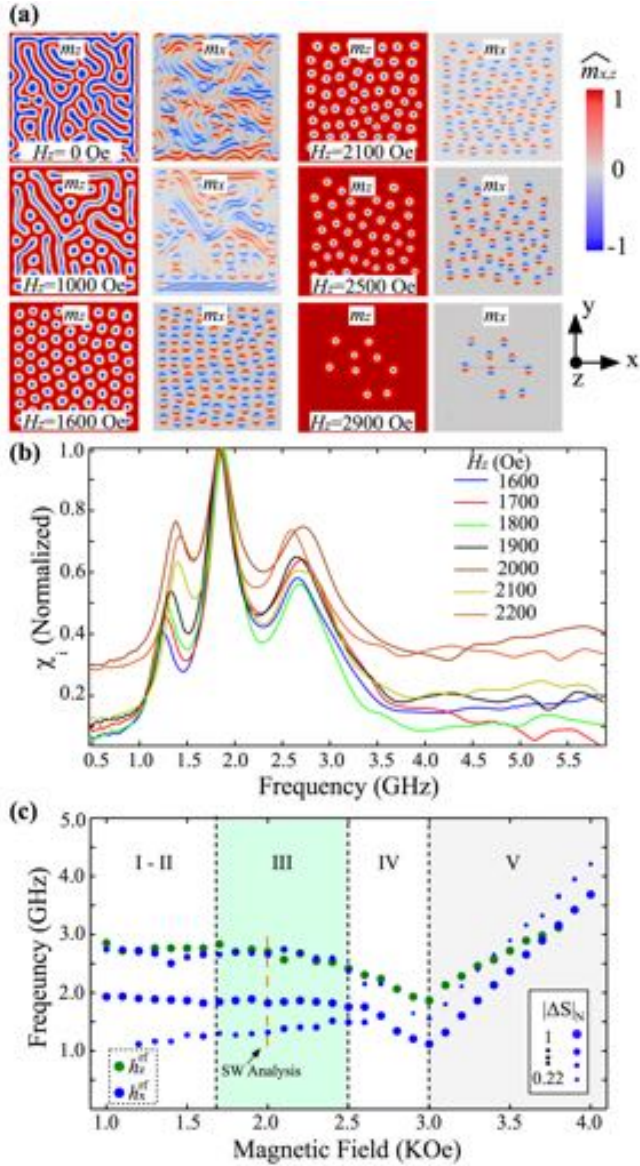

**Figure 5.** Calculated domain morphologies and magnetic susceptibilities. **(a)** The field dependent domain morphology at $H_z$ = 0 Oe, 1200 Oe, 1600 Oe, 2500 Oe and 2900 Oe showing the evolution of disordered stripe domains to ordered skyrmions to disordered skyrmions ($m_z$ at z = 40nm, top side view). The in-plane magnetization along the center of the slab reveals the helicity of these textures ($m_x$ at z = 0nm, top side view). The magnetization distribution ($m_x$, $m_z$) of these equilibrium states is based on the red/blue color-bar. **(b)** The numerically computed susceptibility that results from whole slab at several fixed $H_z$ fields is detailed. **(c)** Fitting the resonance spectra with n-Lorenztian peaks allows us to construct frequency-field *f*-H dispersions. The marker size corresponds to the normalized intensity of the susceptibility at each fixed $H_z$ field and the marker color distinguishes the resonance spectra from either exciting the equilibrium states with an *r.f.* field $h_{rf}$ along the x-axis or the z-axis. Using the equilibrium states, as those in **(a)**, we correlate resonant modes to specific domain textures: (I-II) stripes and coexisting stripes and skyrmions, (III) skyrmion lattice, (IV) disordered skyrmions and (V) Kittel region. These regions are delimited with dashed lines in the *f*-H dispersions.

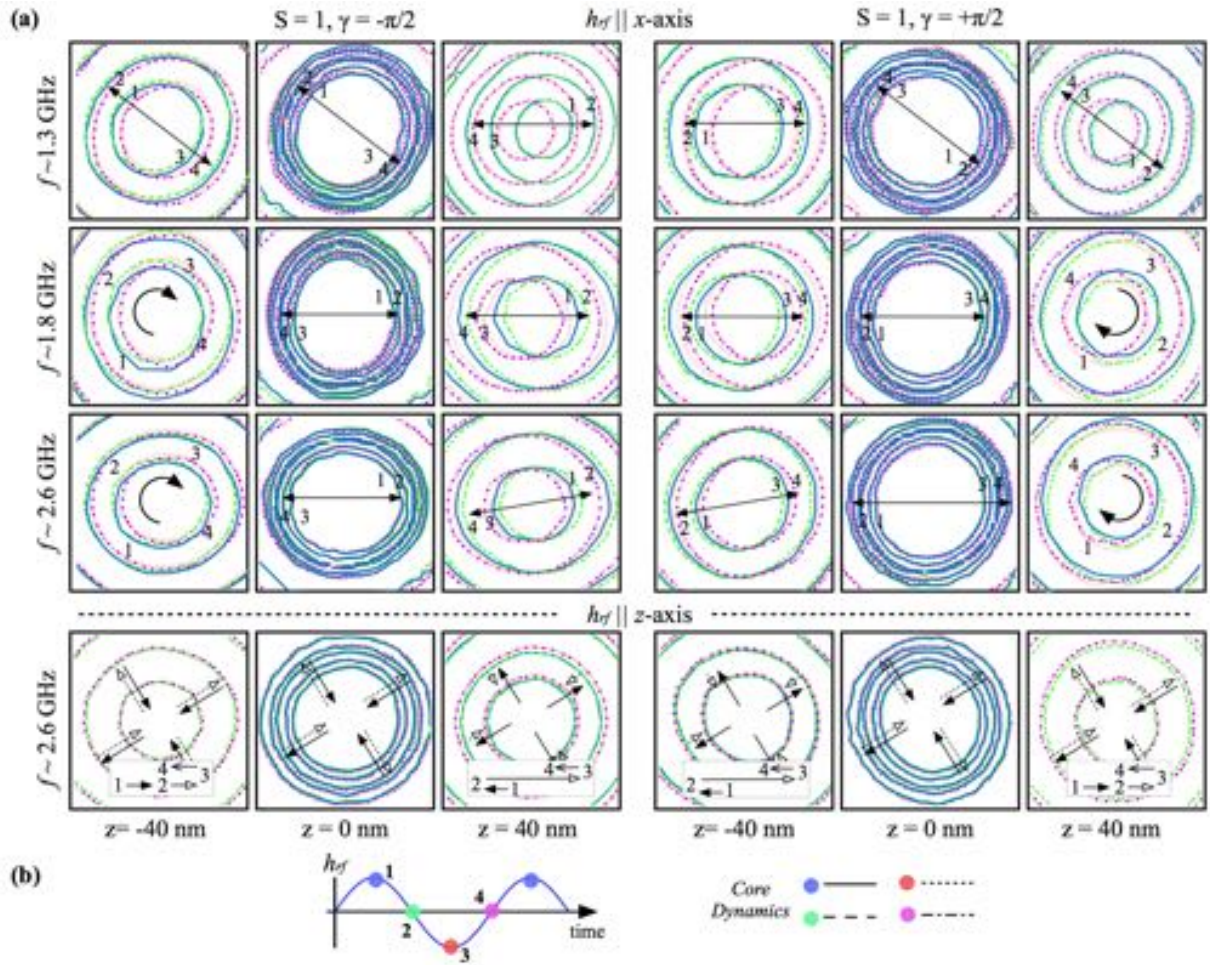

**Figure 6. (a)** The skyrmion localized spin wave dynamics observed from perturbing the skyrmion phase, at $H_z$= 2000 Oe, along both x-axis and z-axis geometries are detailed for both skyrmion helicities. Using contour plots, the motion of the skyrmion is traced along three different depths positions ($z$ = -40 nm, 0 nm, 40 nm) for each resonant mode. Different contour lines depict position of the skyrmion with reference of the **(b)** sinusoidal pulse field and arrows are used as guide for the eye to describe the motion of the spin-wave.

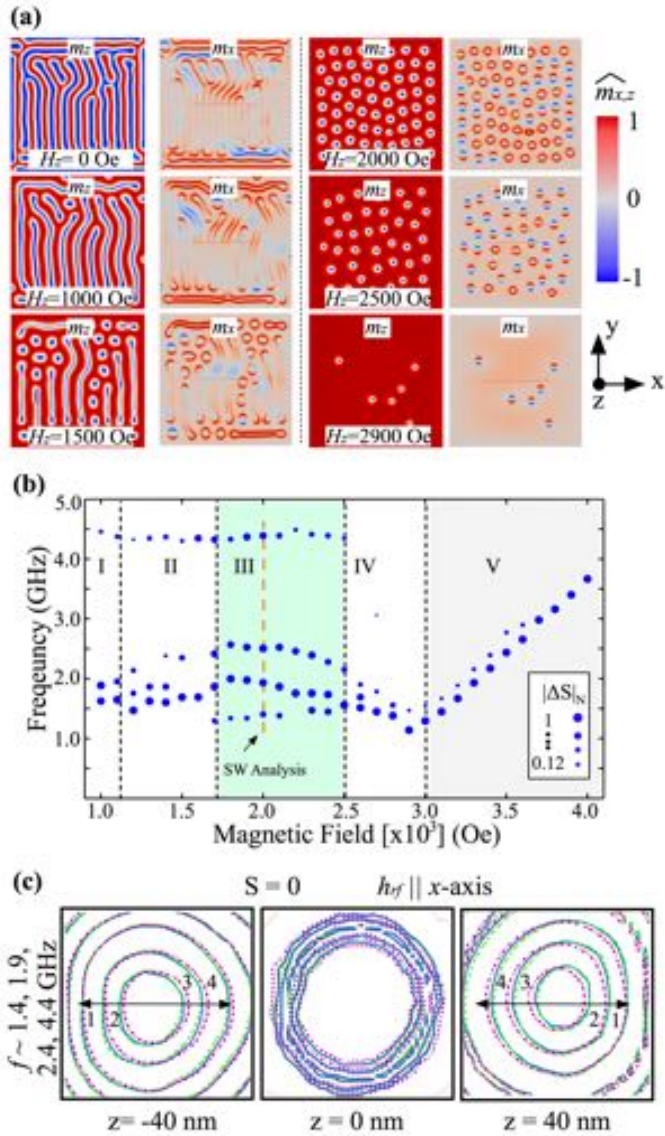

**Figure 7. (a)** The perpendicular field-dependence of equilibrium states with a constant $H_x$ field show the evolution of aligned stripes to cylindrical domains (topside, $z$=40 nm, $m_z$). The cross-section across the center of the slab ($z$=0 nm) shows magnetic texture have a mixture of Bloch-line ordering (top side, $m_x$). **(b)** The computed susceptibility spectra for magnetization states like those in **(a)** result in four mode-branches below magnetic saturation and two-mode branches above magnetic saturation. The magnetization states observed include: (I) stripes, (II) coexisting stripes and cylindrical domains, (III) ordered lattice of cylindrical domains, (IV) disordered cylindrical domains and (V) Kittel region. **(c)** The spin-wave dynamics associated with a bubble domain when probed under an in-plane sinusoidal pulse field.

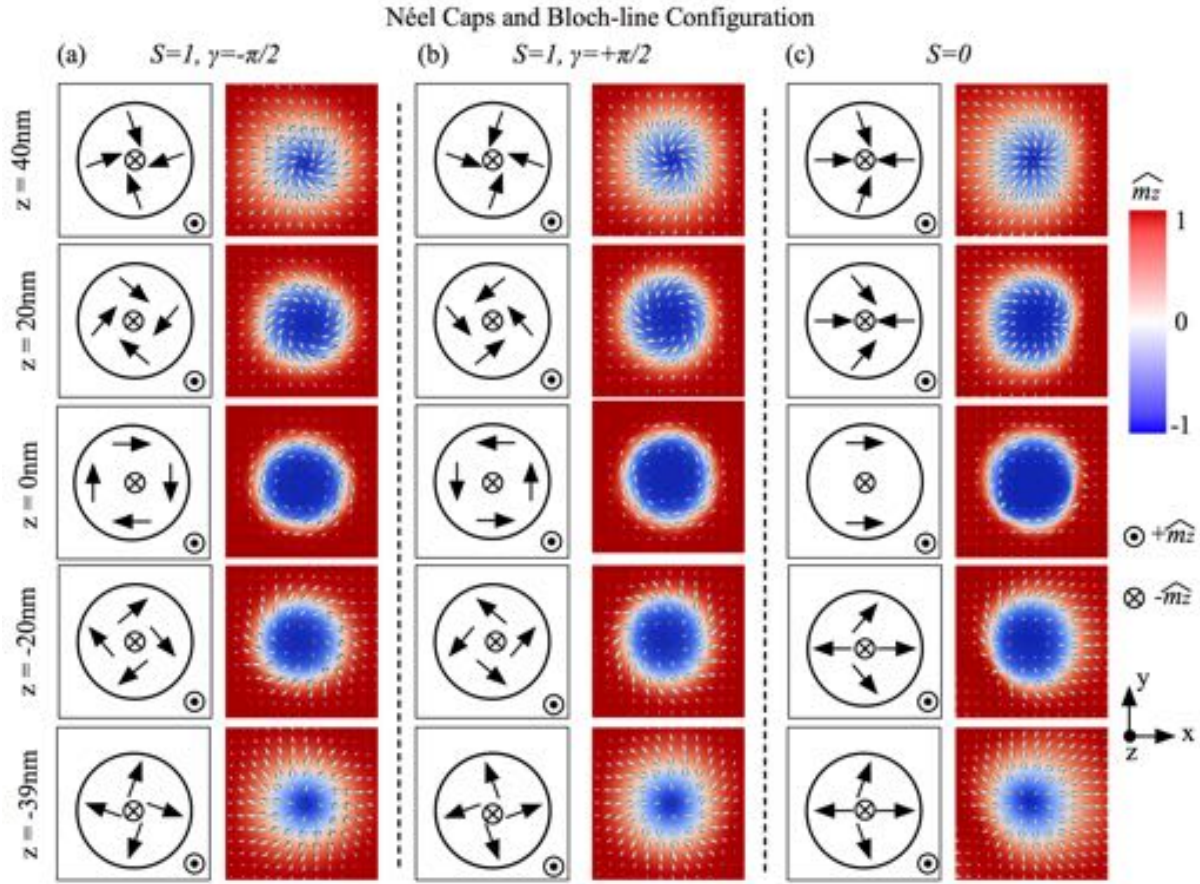

**Figure 8.** The topside-view of the Néel caps and Bloch-line configuration for dipole skyrmions with **(a)** clockwise and **(b)** counterclockwise helicity, and an **(c)** achiral bubble are detailed at different depths (z = 40nm, 20nm, 0nm, -20nm -39nm). The left column shows a simple schematic of the magnetic spin configuration with emphasis on the arrangement of the in-plane magnetic spins (arrows) and the right column shows the domain state obtained from micromagnetic simulations where the white arrows represent the in-plane magnetization and the perpendicular magnetization is represented in terms of the color bar.

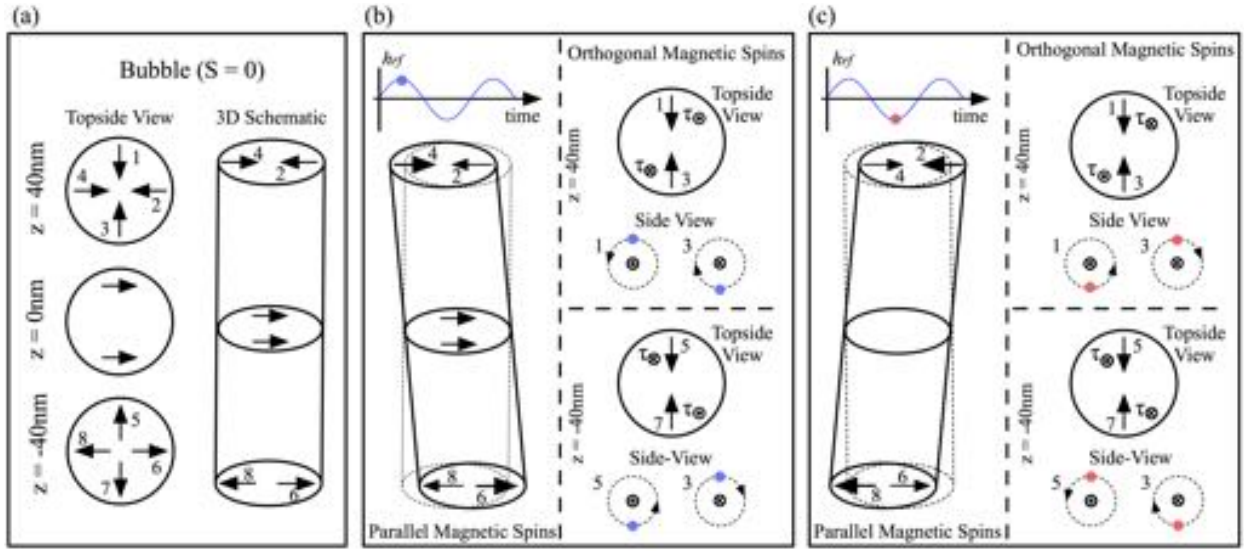

**Figure 9. (a-c)** Schematic of a simplified purely radial achiral bubble domain and its corresponding dynamics under an in-plane microwave field. **(a)** The in-plane magnetization the achiral bubble is emphasized with arrows at three depth positions (z = 40nm, 0nm, -40nm). The 3-D structure of the bubble domain only shows the in-plane magnetic spins along the direction of the perturbation field, and the topside-view shows parallel and orthogonal in-plane magnetic spins. **(b, c)** The redistribution of the magnetic spins are detailed at the maxima **(b)** and minima **(c)** of the sinusoidal perturbation field.

**Supplementary Information**
**Resonant properties of dipole skyrmions in amorphous Fe/Gd multilayers**

S. A. Montoya[1,2], S. Couture[1,2], J. J. Chess[3], J. C. T Lee[3,6], N. Kent[5,6], M.-Y. Im[6,7], S. D. Kevan[3,6], P. Fischer[4,5], B. J. McMorran[3], S. Roy[6], V. Lomakin[1,2], and E.E. Fullerton[1,2 †]

[1]Center for Memory and Recording Research, University of California, San Diego, La Jolla, CA 92093, USA
[2]Department of Electrical and Computer Engineering, University of California, San Diego, La Jolla, CA 92093, USA
[3]Department of Physics, University of Oregon, Eugene OR 97401, USA
[4]Materials Sciences Division, Lawrence Berkeley National Laboratory, Berkeley CA 94720, USA
[5]Physics Department, University of California, Santa Cruz, CA 94056, US
[6]Center for X-ray Optics, Lawrence Berkeley National Laboratory, Berkeley, California 94720, USA
[7]Department of Emerging Materials Science, Daegu Gyeongbuk Institute of Science and Technology, Daegu, Korea

1. **Custom ferromagnetic resonance probe.**

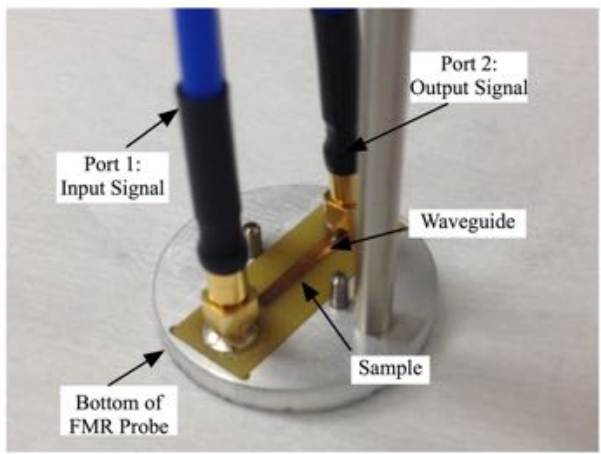

**Supplementary Figure S1. Ferromagnetic Resonance Probe.**

To measure the resonant modes resulting from homogenous and non-homogenous magnetic states in amorphous Fe/Gd films, we built a custom ferromagnetic resonance probe that could be used in combination with a QD PPMS. Supplementary Figure S1 details the bottom portion of the probe where a thin film specimen is fixed on-top of the waveguide. The input/output signals are provided/collected by an Agilent PVNA E8363B.

---

[†] Corresponding author: efullerton@ucsd.edu

## 2. Temperature dependence magnetic hysteresis.

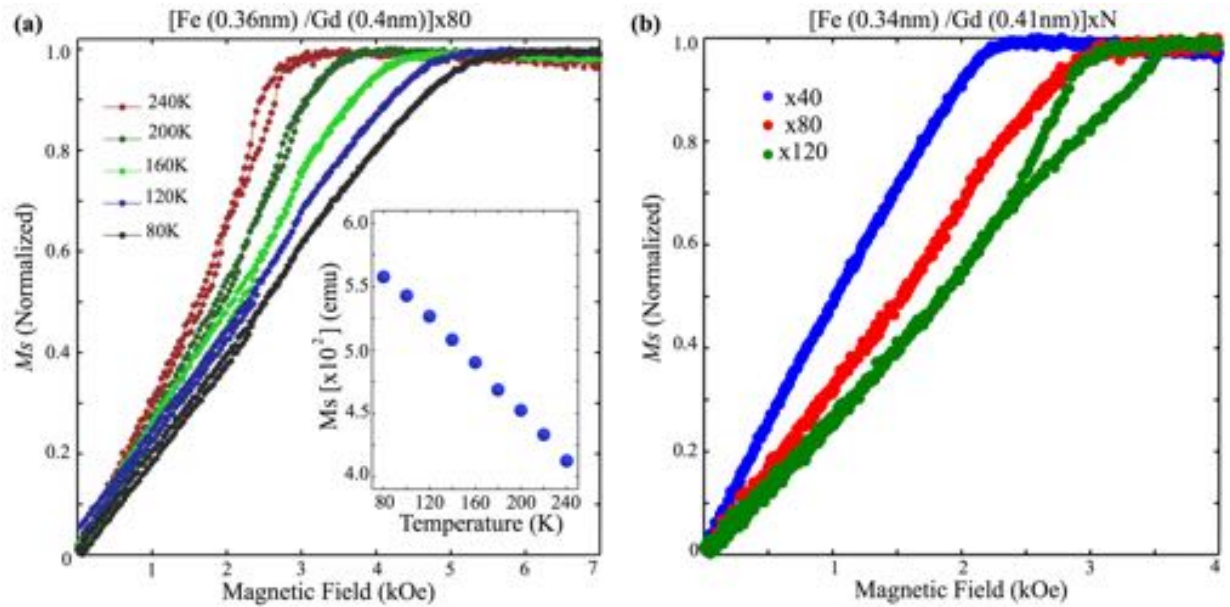

**Supplementary Figure S2.** The magnetic hysteresis for **(a)** [Fe (0.34nm) /Gd (0.4nm)]x80 and **(b)** [Fe (0.34nm) /Gd (0.41nm)]xN where N = 40, 80 and 120 repetitions. The insert in **(a)** shows the moment as a function of temperature for [Fe (0.34nm) /Gd (0.4nm)]x80. The Fe/Gd films are saturated with a field $H_z = \pm 10$ KOe, but we detail smaller magnetic field region.

Supplementary Figure S2 shows normalized magnetic hysteresis loops of the Fe/Gd films presented in this work. For the temperature dependent study, the magnetic loops detail the field-dependence of [Fe(0.36nm) /Gd(0.4nm)]x80 from 240 K to 80 K (Supp. Fig. S2a). The insert of Supp. Figure S2a shows the magnetization as a function of temperature. From 240K to 200K, there is evidence of perpendicular stripe domains in the magnetic loop which can be inferred from the hysteresis around the saturation field. At lower temperatures, stripe domains cannot be inferred from the magnetic hysteresis. For the thickness dependent study, the magnetic loops of [Fe(0.34nm) /Gd(0.41nm)]xN are detailed for three bilayer repetitions (N=40, 80, 120) in Supp. Fig. S2b. At 40 repetitions, the magnetic loops suggest the film prefers to have magnetization lying in the plane of the film. As the number of bilayer are increased, the film exhibits evidence of disordered stripe domains at 80 and 120 repetitions.

### 3. Field effects on resonant spectra.

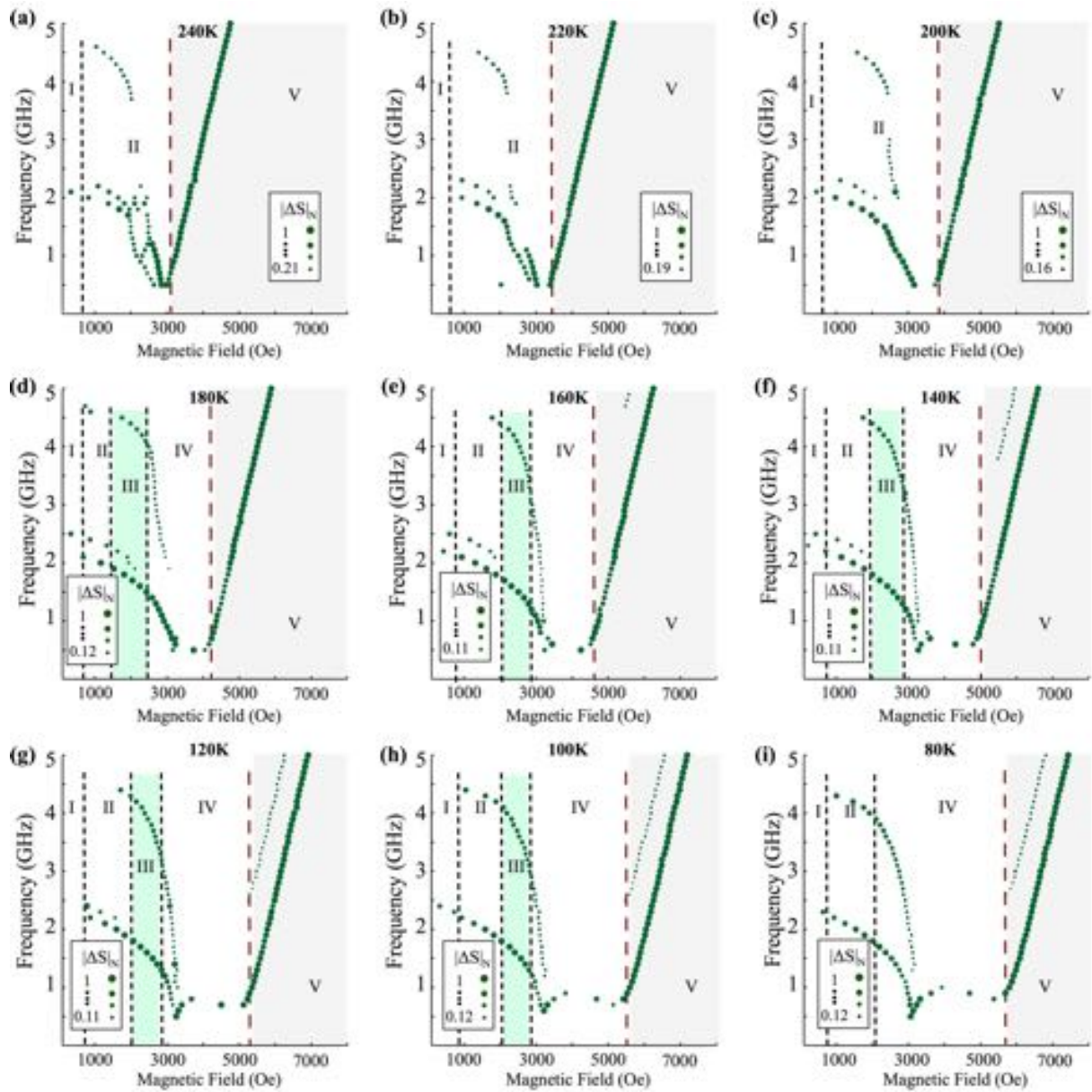

**Supplementary Figure S3.** Temperature dependence of the resonant modes of a [Fe (0.36nm) / Gd (0.40nm)]x80 multilayer under a different field history to Fig. 3 (main text). The frequency-field *f-H* dispersions from 240 K to 80 K were constructed from measurements were the perpendicular magnetic field is swept from $H_z$= 0 Oe to 10 kOe at each fixed frequency. At each temperature, several regions are identified that correspond to different magnetization states: (I) stripes, (II) coexisting stripes and skyrmions, (III) skyrmion lattice, (IV) disordered skyrmions, and (V) Kittel region. These regions are delimited with dashed lines in the *f-H* dispersions. The saturation field $H_{sat}$ is obtained from magnetic hysteresis measurements performed in the same geometry.

The field dependence of the resonant spectra was also investigated, by first magnetically saturating the sample in a positive field, then reducing the magnetic field to zero-field before scanning the resonance spectra from remanence to positive saturation (Supp. Fig. S3). We find the resonant mode branches are very similar to those in Fig. 3 (main text) when scanned from negative to positive magnetic saturation, with the exception of an additional mode-branch appearing around $f \sim 2.5$ GHz at temperature spanning from 180 K to 120K (Supp. Figs. S3d-g). The presence of this additional mode-branch indicates a variation in the domain morphology as a result of the different field history. As the temperature is reduced from 180 K to 100 K we further observe that the mode-branch appearing at $f \sim 2.3$ GHz shifts upwards in frequency toward the mode-branch at $f \sim 2.5$ GHz until a single resonant branch is observable (Supp. Fig. S3d-h).

## 4.  Néel cap configuration for cylindrical domains with different chirality.

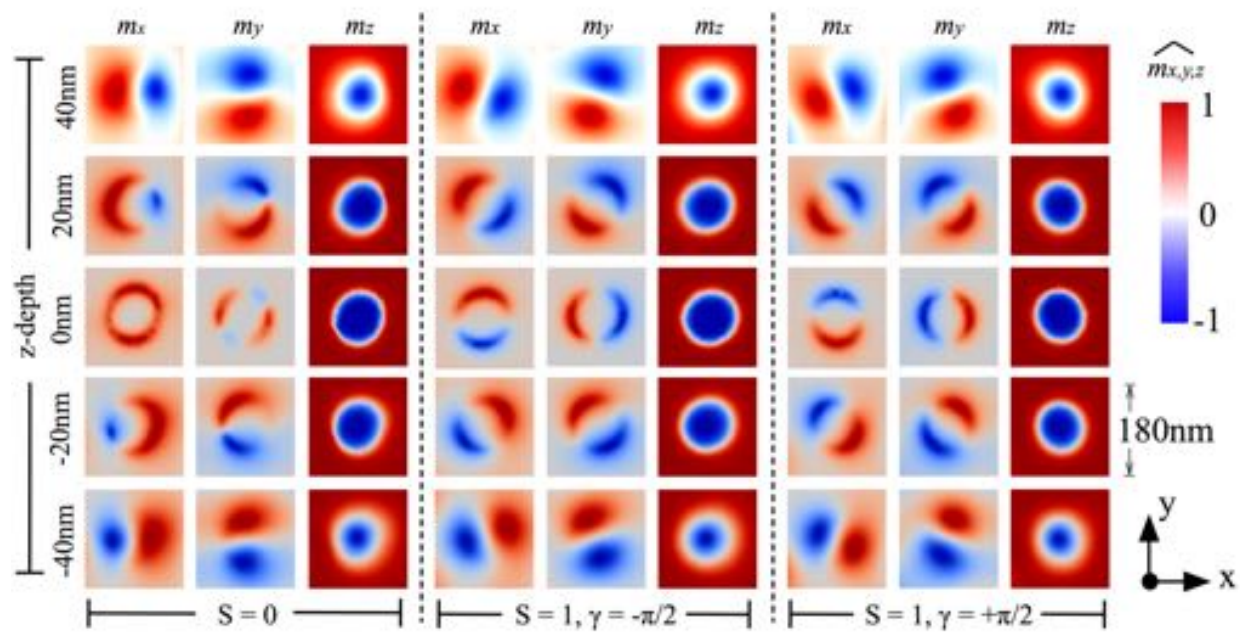

**Supplementary Figure S4.** The domain morphology of cylindrical domains with different chirality (S = 0, 1) are explored at different depths of the slab and magnetization ($m_x$, $m_y$, $m_z$) orientations. Each equilibrium state depicts the top-side view of 180nm$^2$ area.

The domain morphology across the thickness of the slab for cylindrical domains with different chirality is detailed. Both dipole skyrmions (S = 1, $\gamma = \pm\pi/2$) and bubbles (S = 0) show similar perpendicular magnetization ($m_z$) distribution; however, the lateral magnetization ($m_x$, $m_y$) varies throughout the thickness of the slab for all the textures. Differences are observed in the Bloch-line arrangement (center of the slab, z = 0nm) and the Néel cap configuration (away from the center of the slab). In the case of bubbles, we see the Bloch-line aligns in the direction of the applied in-plane field. We recall, that the observation of bubbles in these films results from perturbing the equilibrium states with a constant in-plane field along the x-direction, with a magnitude $H_x = 60$ Oe, while a perpendicular field $H_z$ is applied. Effects of the in-plane field on the domain morphology is detailed in the Supplementary Section 7. In contrast, skyrmions (S = 1, $\gamma = \pm \pi/2$) have random chiral Bloch-line order which is set by the minimization of the magnetic energies in the system.

The most distinct difference is observed in the Néel cap configuration surrounding the cylindrical domain. For bubbles (S = 0), the Néel caps orientation along both $m_x$ and $m_y$ are mirror images around the center of the slab. This is the common configuration expected for film with a Bloch domain structure. For skyrmions (S=1, $\gamma=\pm \pi/2$), the Néel cap arrangement depends on the chirality and the mirror images of both $m_x$ and $m_y$ are winded by ±90 degrees from the top to the bottom of the slab. The chiral domain wall is what provides dipole skyrmions their topological protection.

## 5. Spin-wave resonance above magnetic saturation.

Using micromagnetic simulations, we explored the magnetization dynamics that result in the second mode branch appearing above magnetic saturation in the numerically computed susceptibility (Fig. 5c, main text). We apply a sinusoidal field pulse along the x-direction to a magnetization state that is uniformly magnetized along the direction of the perpendicular field ($+H_z$) and exhibits a secondary resonant mode at $H_z = 3500$ Oe with resonance at $f \sim 2.89$ GHz (Fig. 5c, main text). Inspecting the dynamics across the slab reveals a spin wave originates from the edge of the slab and moves across the thickness of the slab in the direction of the field perturbation (Supp. Fig. S5). The magnetization state at time interval $t_1$ shows $m_z$ at different depths of the slab (z = 40 nm, 20 nm, 0 nm, -20 nm, -40 nm). Here, contour lines depict the regions of the magnetization that are not completely aligned in the direction of the perpendicular magnetic field (Supp. Fig S5a). The highest deviations are observed along the top and bottom edges of the slab and appear negligible around the center of the slab. One can further observe the deviations from the top and bottom of the slab along $m_z$ are at opposite sides. We further explore the spin-wave moving through the slab by plotting variations of $m_x$ at the same depth positions as $m_z$ for several time intervals ($t_1$, $t_2$, $t_3$, $t_4$) detailing the sinusoidal pulse (Supp. Fig. S5b). This representation allows us to further visualize the spin wave originating from the edge of the slab, which ultimately results in the additional weak mode above magnetic saturation.

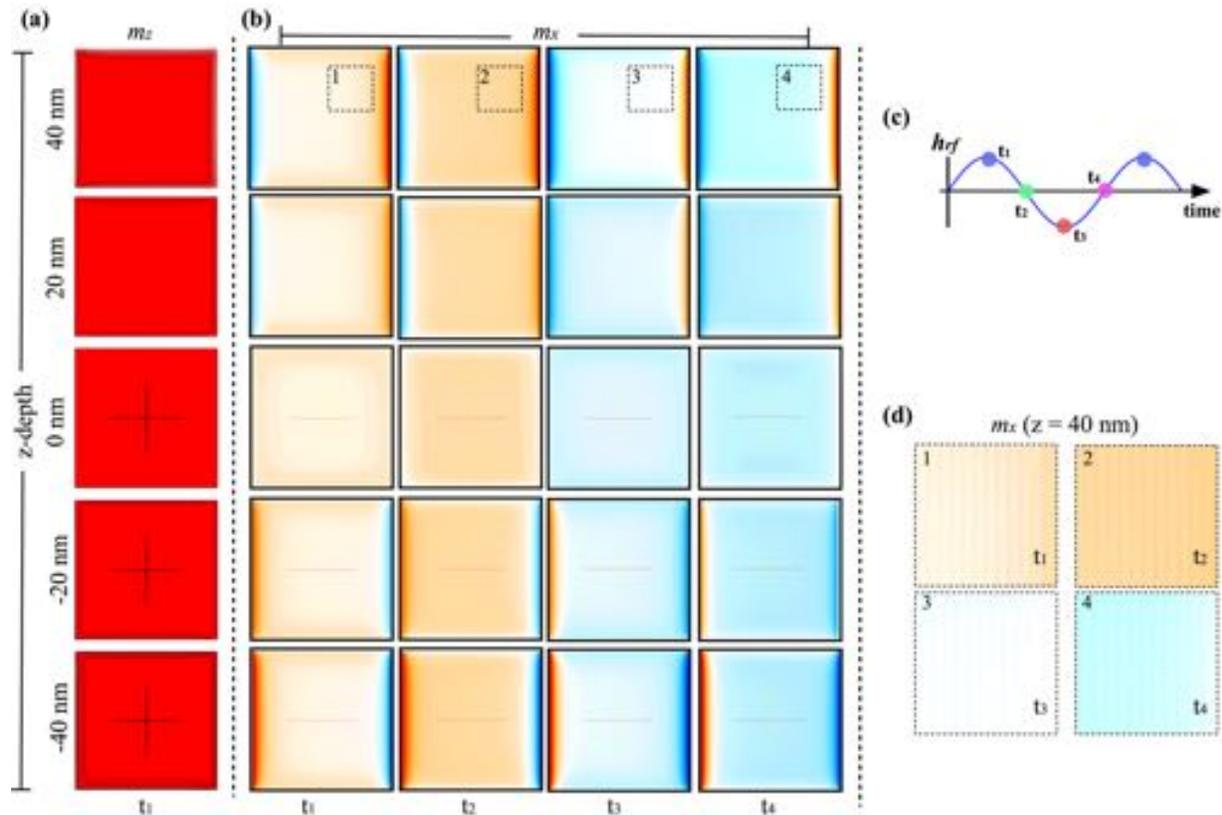

**Supplementary Figure S5.** **(a, b)** The top-side magnetic distribution of $M_x$ and $M_z$ are detailed as a function of slab thickness and evolution in time when perturbed with a small oscillating $h_{rf}$ in-plane field. **(c)** The $h_{rf}$ field is applied along the x-direction, where several snapshots at different time intervals are collected. Here we detail four snapshots ($t_1$, $t_2$, $t_3$, $t_4$) that show a spin wave moving through the thickness of the slab. The cross at the center of the ES is an artifact from the visualization software. **(d)** Enlarged regions from time distributions along $m_x$ show a moving spin wave.

## 6. Effects on the susceptibility resulting from exchange length variations.

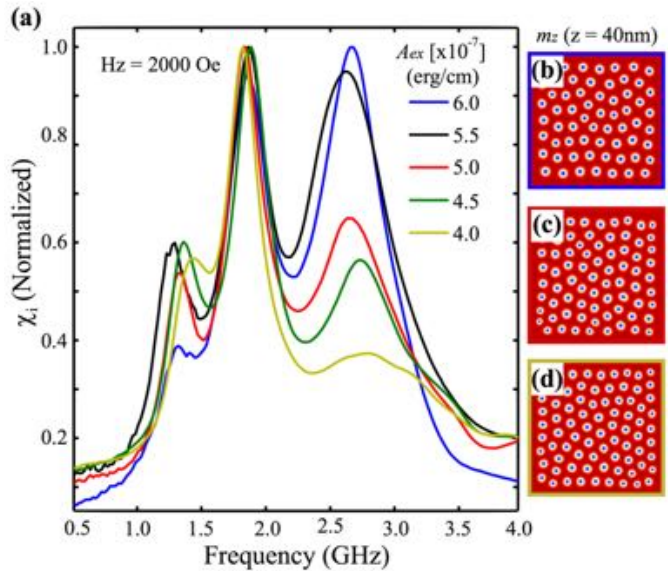

**Supplementary Figure S6. (a)** The susceptibility for different exchange A strength is computed for a fixed field $H_z$ = 2000 Oe. The top-side magnetization states detail $m_z$ at the top of the slab (z = 40 nm) under $H_z$ = 2000 Oe for different exchange: **(b)** $A$ = 6x10$^{-7}$ erg/cm, **(c)** $A$ = 5x10$^{-7}$ erg/cm and **(d)** $A$ = 4x10$^{-7}$ erg/cm.

As we have detailed in Ref. [2], the exchange coupling $A$ plays an important role in the skyrmion long range order, as well as, the size of the skyrmion, volume fraction of perpendicular domain and Néel caps, etc. To investigate how variations of exchange length $A$ affect the skyrmion resonant properties, we varied the $A$ strength from 4 to 6x10$^{-7}$ erg/cm, fixing $M_S$ = 400 emu/cm$^3$ and $K_U$ = 4x10$^5$ erg/cm$^3$, and modeled the susceptibility from resulting from the equilibrium states at $H_z$= 2000 Oe (Supp. Fig. S6a). All these exchange coupling values result in an ordered skyrmion phase (Supp. Fig. S6b-d) with resonances appearing at $f \sim$ 1.3GHz, 1.8G Hz and 2.6 GHz. As the exchange coupling is increased the absorption intensity solely varies at resonances elicited at $f \sim$ 1.3 GHz and 2.6 GHz, which we know are related to the skyrmion domain (Fig. 5, main text). Interestingly, these resonate at different maximum/minima intensities depending on the exchange coupling strength. In the case of $A$ = 6x10$^{-7}$ erg/cm, the two primary resonances appear at $f \sim$ 1.8 GHz and 2.6 GHz; whereas, for $A$ = 4x10$^{-7}$ erg/cm the primary resonances appear at $f \sim$ 1.3GHz and 1.8 GHz. This evidence suggests that our inability to experimentally detect the mode-branch appearing around $f \sim$ 1.3 GHz in the Fe/Gd films is likely a result of either: (i) the exchange coupling is higher than numerically assumed or (ii) the intensity absorption at this resonance is weak enough that we are unable to discern it.

### 7.  Local susceptibility from skyrmion domain.

To further probe the spin-waves, we performed localized susceptibility modeling on the skyrmion lattice phase when $H_z = 2000$ Oe, by analyzing local volume regions that either contains a single skyrmion or an area between skyrmions as shown in Supp. Fig. S7a and then compared it to the susceptibility averaged over the whole slab. The different regions examined result in similar resonance spectra (Supp. Fig. S7b). but with some differences in normalized peak amplitude – the skyrmion domains have much stronger fluctuations at resonances occurring at $f \sim 1.3$ GHz and 2.6 GHz, in contrast, a region without a skyrmion resonates strongest at $f \sim 1.8$ GHz. Overall the localized susceptibility indicates the whole slab containing the non-homogenous magnetic textures resonates at the same frequencies and no single magnetic features drives the skyrmion localized spin-wave modes.

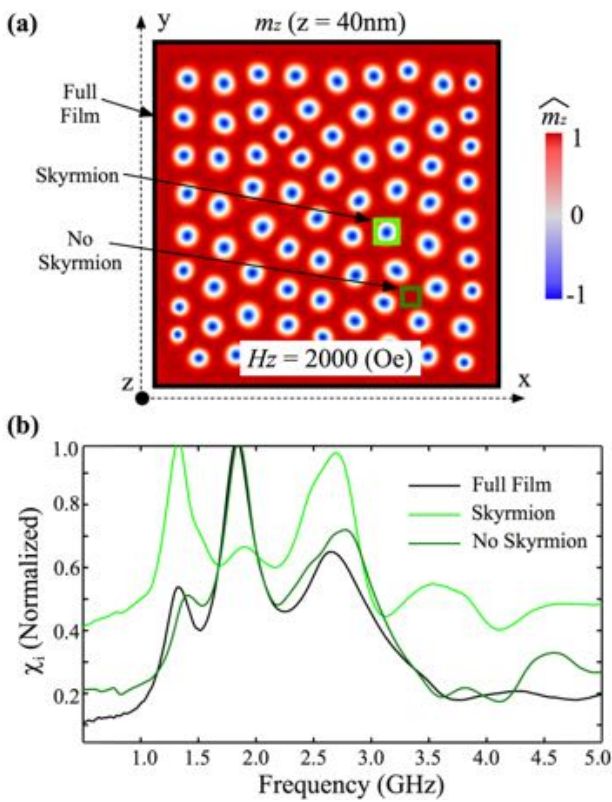

**Supplementary Figure S7.** The susceptibility contribution of specific regions of the skyrmion phase stabilized at $H_z = 2000$ Oe are compared to investigate the source of the resonance spectra. The analyzed areas are shown on the top-side view of the domain morphology with contributions along $m_z$ **(a)** and $m_x$ **(b)**. The regions highlight a single skyrmion, no skyrmion domain and the resonance spectra is compared to that of the whole film. **(c)** The complex susceptibility ($\chi_i$) reveals the three different areas resonate at the same frequency with variations in absorption intensity.

## 8. Bloch-line and Néel caps configuration under external in-plane field.

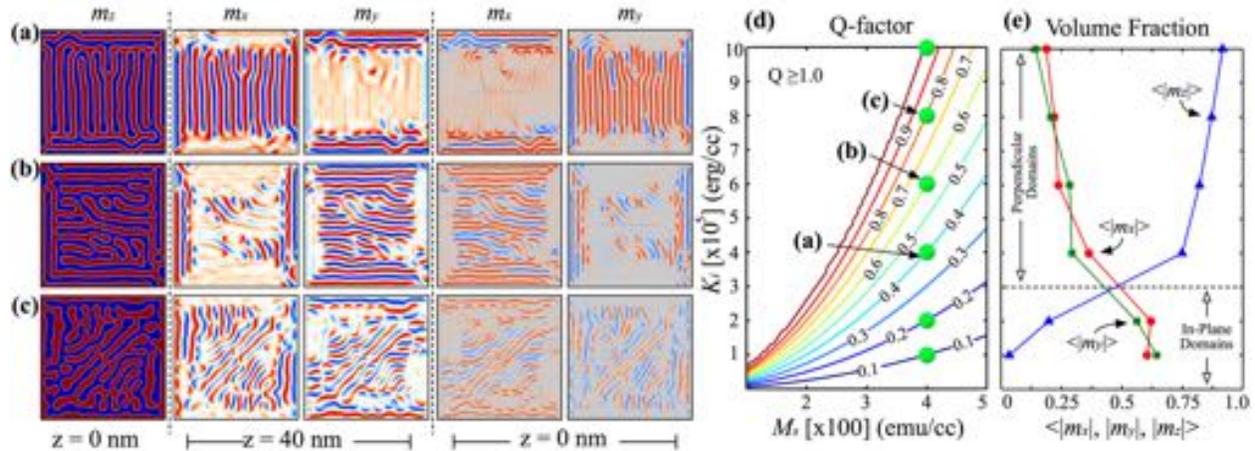

**Supplementary Figure S8.** The rearrangement of equilibrium states with different $Q$-factor ratios under an external in-plane field, along the x-axis, are detailed. **(a-c)** The top side view ordering of $m_x$ and $m_y$ at the top ($z = 40nm$) and center ($z = 0nm$) of the slab reveals the rearrangement of the Néel caps and Bloch-line. **(e)** The magnetization volume fraction $<|m_{x, y, z}|>$ for the equilibrium states in **(d)** show that perpendicular magnetic domains with Néel caps become favorable when $<|m_z|>$ is greater than $<|m_{x, y}|>$. In this regime, as the $Q$-factor increases the volume fraction of the in-plane domains decreases, whereas, perpendicular domains become more prevalent in the domain morphology.

To understand the arrangement of the Bloch-line under an external in-plane field, we evaluated the domain morphology that results from different ratios of anisotropy $K_U$ and magneto-static energy $2\pi M_S^2$, which is defined as the material $Q$-factor [1]. It follows that thin films with $Q$-factor < 1 will prefer to have the magnetization lying on the plane of film; whereas, magnetic specimens with $Q$-factor > 1 will have the magnetization aligned perpendicular to film. However, in the case of relatively thick films with material $Q$-factor < 1, the minimization of anisotropy and magneto-static energy will favor a domain structure consisting of perpendicular magnetic domains and closure domains, e.g. Néel caps. As detailed in Ref. [2], our observation of perpendicular domains in these Fe/Gd films is a result of the latter circumstance.

Our sampling of magnetization states with varying $Q$-factor (~0.1 to 1) allows us to capture domains structures with different ratios of perpendicular domains and Néel caps (Supp. Fig. S8). At zero-field, we find that magnetization states with $K_U \leq 3x10^5$ erg/cm$^3$ favor a vortex domain state given the area restrictions of the slab, i.e. the magnetization would be in-plane for a continuous thin film; whereas equilibrium states with $K_U > 3x10^5$ erg/cm$^3$ favor perpendicular disordered stripe magnetic domains. When these magnetization states are exposed to an in-plane field along the x-direction, with a magnitude of $H_x = 60$ Oe, the following broad observations can be made:

    **i.** $K_U = 4\text{x}10^5$ erg/cm$^3$ – The majority of Néel caps align in the direction of the in-plane field ($m_x$, $z = 40$ nm) and the Bloch-line of most stripe domains points orthogonal to the the $H_x$ field ($m_y$, $z = 0$ nm).

    **ii.** $K_U = 6\text{x}10^5$ erg/cm$^3$ – The Bloch-line of most stripe domains align in the direction of the in-plane field ($m_x$, $z = 0$ nm) and the Néel caps are orthogonal to the $H_x$ field ($m_x$, $z = 40$ nm).

    **iii.** $K_U \geq 8\text{x}10^5$ erg/cm$^3$ – The in-plane field $H_x$ is insufficient to align a specific lateral magnetization component across the film. At the top of the slab, the Néel caps have magnetization randomly distributed along both $m_x$ and $m_y$ ($z = 40$ nm); similar observations can be reached for the Bloch-line across the center of the slab ($z = 0$ nm).

In all the cases above, we observe a transverse alignment of the Bloch-line and Néel caps because these structures have a helical arrangement across the thickness of the slab. Depending on the volume fraction of the in-plane magnetic domains, we observe distinct structures align differently with the in-plane field (Supp. Fig. S8e). As the volume fraction of the domain wall (Néel caps and Bloch-line) decreases, we find the in-plane field $H_x$ is unable to reorder the domain state. Altogether, it appears that a specific ratio between $<|m_{x,\,y}|>$ and $<|m_z|>$ is required to favor the of the alignment of the Néel caps or the Bloch-line with an external field.

From these simulations, we can infer that the presence of an in-plane can result in helicity changes for domain states at the ground state. Effects on the chirality of the domains at zero-field will affect the perpendicular magnetic domains that form under an out-of-plane magnetic field. In our case, bubbles result from aligning the stripes at the ground state (Fig. 7a, main text).